\documentclass[10pt, amsmath, amssymb, twocolumn]{revtex4-2}
\UseRawInputEncoding
\usepackage[normalem]{ulem}
\usepackage{url}
\usepackage[colorlinks=true, allcolors=blue]{hyperref}
\usepackage{graphicx}

\usepackage{bm}
\usepackage[utf8]{inputenc}
\usepackage[T1]{fontenc}
\usepackage{mathptmx}
\usepackage{etoolbox}

\makeatletter
\def\@email#1#2{
 \endgroup
 \patchcmd{\titleblock@produce}
  {\frontmatter@RRAPformat}
  {\frontmatter@RRAPformat{\produce@RRAP{*#1\href{mailto:#2}{#2}}}\frontmatter@RRAPformat}
  {}{}
}
\makeatother
\begin{document}

\title{Topological photonics in nanoscaled systems with far field radiation and polarization singularities.}

\author{G. Salerno}
 \affiliation{Department of Applied Physics, Aalto University School of Science, P.O. Box 15100, Aalto FI-00076, Finland}
\affiliation{Dipartimento di Fisica ``E. Fermi'', Università di Pisa, Largo Bruno Pontecorvo 3, Pisa, 56127, Italy}
\email{grazia.salerno@unipi.it}
\date{\today}

\begin{abstract}
Topology is a powerful framework for controlling and manipulating light, minimizing detrimental perturbations on the photonic properties. Combining nanophotonics with topological concepts presents opportunities for both fundamental physics and technological applications. Although most topological photonic realizations have been inspired by condensed-matter analogue models, new topological ideas have just begun to be realized at the nanoscale. Nanophotonics is characterized by subtle phenomena that are not usually considered in other topological models' realizations, such as nonlocality, strong field confinement, and light radiating to the far-field continuum. In this perspective, we will discuss how standard topological band theory for photonic crystals needs to be extended by a more comprehensive approach that properly treats such nanophotonic intrinsic effects and, in particular, the interplay of polarization and far-field radiation. We highlight the emerging role that polarization singularities might play in defining the topological invariants in the far field, which are not fully captured by bulk observables alone. 
We conclude by outlining a set of open questions and promising directions for exploring novel concepts in topological nanophotonics and shaping next-generation photonic devices.
\end{abstract}

\maketitle

\section{Introduction}

The ability to engineer matter at the nanoscale is a research field that has revolutionized science and technology.
Nanophotonic systems, such as photonic crystal metasurfaces, nano-ring resonators, or plasmonic nanoparticles, are structures fabricated on the scale of light's wavelength with strongly confining high-index materials. These systems offer unprecedented ways of controlling light at room temperature with remarkable features, from ultrafast dynamics in the order of femtoseconds and chip-scale integration to band dispersion control and light confinement into volumes beyond the diffraction limit~\cite{koenderink2015nanophotonics, WangPaivi_rev, wang2018nanophotonic}. The extreme spatial confinement, for example, boosts the electromagnetic field, significantly enhancing light–matter interactions, lowering the pump power requirements~\cite{Kravets}. All of these are crucial ingredients for reaching the few-photon limit beyond nonlinear effects, thus making nanophotonics an ideal platform for on-chip compact devices useful in quantum information technologies~\cite{karabchevsky2020chip, gonzalez2024light}. 

Controlling photons while avoiding detrimental effects on their properties is one of the goals for quantum optical applications~\cite{yan2021quantum}. Topology, an area of mathematics that concerns geometrical properties immune to deformations, emerges as a promising research avenue to protect systems from disorder and imperfections. Topology was first introduced in condensed matter physics to explain the quantum Hall effect in topological insulators~\cite{Yoshioka}, for which the current takes integer values multiple of a ``quantum of current''. This quantized current is restricted to the edges of the system and is unidirectional, protected from back scattering despite impurities in the material. These effects are captured by topological invariants described by the topological band theory, and are not limited to solid-state materials, but have been realized in many systems, including photons on a lattice~\cite{Ozawa_Rev, Price2022roadmap}.

A topological phase transition happens when the band gap closes and reopens, during which the bands become degenerate and the wavefunctions undergo a singular change in their topology, similar to piercing a sphere to transform it into a torus. Dirac points are a fundamental type of band degeneracy, around which topological properties, such as non-zero Berry phases, are often found. For this reason, Dirac-like points, including Weyl points in 3D and exceptional points in non-Hermitian systems with gain and loss~\cite{li2023exceptional}, have received great attention in photonics as a playground for topological phenomena~\cite{weick2013dirac, lu2014topological}. In combination with strain design techniques borrowed from the solid-state graphene community~\cite{goerbig2011electronic}, synthetic pseudomagnetic fields can be engineered around such Dirac singularities, and Landau quantization can be directly observed~\cite{Salerno2020c, mann2020tunable, barczyk2024observation}.

Chiral effects are another interesting aspect of topological implementations. Breaking in‑plane symmetries with chiral structures might lift mode degeneracies and generate non-zero local Berry curvatures in the photonic bands~\cite{solnyshkov2016chirality, han2024chiral}. Various chiral nanophotonic metasurfaces have been implemented, showing superchiral near‑fields and anomalous polarization behavior across the band diagram~\cite{zanotto2019photonic}, and circular dichroism~\cite{ali2023circular, ali2024maximum}. When paired with an explicit time‐reversal–symmetry‐breaking mechanism, chiral phenomena can show the non‐reciprocal behavior essential for robust topological nanophotonic devices~\cite{raghu2008analogs, haldane2008possible, wang2008reflection}. For example, magnetic nanoparticles have been used in plasmonics to achieve lasing in chiral modes activated by an external magnetic field~\cite{freire2022magnetic}. Magnetic effects such as Zeeman splitting of an exciton-polariton system can also provide Chern bands beyond Dirac points~\cite{nalitov2015polariton, xie2025polariton}. However, for practical and scalable implementations, it is highly desirable to engineer topological phenomena without relying on external magnetic fields. In this context, artificial gauge fields are a central ingredient in many photonic topological models, particularly those inspired by solid-state analogues~\cite{Ozawa_Rev}.

Work on topological nanophotonics has considered lattices that directly map onto paradigmatic tight-binding models. Notable realizations include dimerized 1D chains realizing the Su-Schrieffer-Heeger model~\cite{LingSSH, Pocock}, 1D quasiperiodic chains implementing the Aubry-Andrè-Harper model~\cite{poshakinskiy2014radiative}, the Kagome lattice~\cite{proctor2021higher} or shrunken and expanded honeycomb arrays corresponding to the Wu-Hu model~\cite{siroki2017topological, gorlach2018far, honari2019topological, WuHu, proctor2020robustness}. Rather than relying solely on coupling between well-separated particles, connecting them with waveguides of various widths has been demonstrated to be an effective way to modulate the couplings via near-field, although with reduced strength due to screening~\cite{yan2025near}. 

While waveguides and ring resonators are a versatile platform for exploring topological photonic effects~\cite{hafezi_robust_2011, hafezi_imaging_2013, leykam2020topological}, from synthetic dimensions~\cite{Fan_synthetic_rev, SegevSynthetic} to topological lasing~\cite{SegevLasing2, SegevLasing}, the primary focus of this Perspective lies on photonic crystals where far-field radiation and polarization singularities emerge as key ingredients for topological phenomena.
When long-range, radiative and retardation effects are included, the simple Hermitian tight-binding picture is not applicable anymore. Unintended radiation leakage is negligible in regular optics due to large mode volumes, but it dominates at the nanoscale, where it is often not easily adjustable, occurring at a wide angular and spectral range. Radiative effects can strongly constrain the quality factor of the modes, degrading nonlinear interaction strength, and broadening spectral lines.
As a result, the topological character of these open photonic platforms can be modified, and sometimes even enriched~\cite{Pocock, poshakinskiy2014radiative, wong2024classifying}. Embracing both the intrinsically non-Hermitian nature of nanoscale photonic lattices and their nonlocality is therefore crucial for accurately defining the topology of light’s radiation fields~\cite{RiderRev_Perspective, monticone2025roadmap}. 

\subsection*{Bound states in the continuum as topological defects}
In photonic crystals, light can either be localized within the structure or have modes that lie above the light line and radiate into the far field. These latter modes, known as \textit{guided resonances}, are subject to radiation losses and correspond to out-of-plane diffraction of Bloch waves and thus can carry information about the bulk to the far field. 

Among guided resonances, bound states in the continuum (BICs) have recently emerged as an invigorating addition to topological phenomena of light~\cite{hwang2022reviewBIC, SoljacicTopoBic}. BICs are localized modes embedded in dispersive bands appearing as light polarization singularities decoupled from the free-space radiation~\cite{bliokh2019geometric, KivsharRev}. BICs are protected from radiation decay by the existence of a quantized topological charge that tells how often the light polarization vector winds around the BIC’s momentum, namely its vortex center~\cite{SoljacicTopoBic, Alu}. Due to the impossibility of defining a polarization vector at the BIC core, these states appear dark in the far field.
The BIC topological charge can be calculated as the winding number of the polarization vector~\cite{SoljacicTopoBic}
\begin{equation}
    q = \frac{1}{2\pi} \oint_\mathcal{C} \nabla_\mathbf{k} \phi(\mathbf{k}) \cdot d \mathbf{k} ,
    \label{topocharge_def}
\end{equation} 
where $\phi(\mathbf{k})$ is the phase of the linear polarization vector. The sign of the topological charge gives the orientation of the winding. 

The topological protection of BICs at high-symmetry points of the Brillouin zone also stems from the rotational symmetry of the real space lattice. The topological charge $q$ of a symmetry-protected mode that transforms according to a specific irreducible representation of the system’s point group can be derived as~\cite{Salerno2022, arjas2024high}
\begin{equation}
    q_{\text{irrep}} =  1 - \frac{n}{2\pi}\arg(\epsilon_\text{irrep}),
    \label{topocharge}
\end{equation}
where $\epsilon_\text{irrep}$ is the character of that representation under a $2\pi/n$ rotation and is taken from the $C_n$ character table. This relation implies that point groups with higher rotational order $n$ can support symmetry-protected BIC modes with larger quantized topological charges, which can be realized in quasicrystalline structures~\cite{arjas2024high}.

BICs are observed in photonic crystals with nearly infinite lifetimes and extremely high quality factors. For these reasons, they are particularly suited for low-threshold lasing~\cite{KodigalaLasingBIC, Heilmann2022}, and guiding slow light~\cite{tanimura2025slow}. BICs might also be useful for enhancing strong nonlinear responses~\cite{carletti2018giant, krasikov2018nonlinear, dang2022realization}, achieving Bose-Einstein condensation of exciton-polaritons~\cite{ArdizzoneBIC, efthymiou2024condensation}, and providing quantum information storage~\cite{depaz2023bound}. 
BIC lasing on a nearly-flat band was also observed in dielectric plasmonic arrays with guided modes~\cite{eyvazi2025flat}, thus achieving topological features in combination with a broad angular emission distribution from photonic structures. 

In addition to the symmetry-protected linear polarization vortexes, there can also exist other types of BICs. The so-called accidental, and Friedrich-Wintgen, can appear at any point in the Brillouin zone due to destructive interference~\cite{yang2014analytical, chern2023bound}, from multipolar origin~\cite{sadrieva}, or spawning from bigger BICs~\cite{kang2021merging, le2024super}.
Moreover, symmetry-protected BICs can also break into pairs of circularly-polarized states, with half charges and opposite handness~\cite{yoda2020generation, liu2019circularly, ye2020singular, wang2022realizing, zagaglia2023polarization}.

Despite their topological origin~\cite{SoljacicTopoBic}, BICs are defined in momentum space as isolated polarization singularities, or local \textit{topological defects}. Their connection to the broader framework of topological band theory, based on global invariants and bulk-boundary correspondence in open radiative systems, is only beginning to be clarified.
To unveil this connection, we now turn to topological band theory as applied to nanophotonic lattices.

\section{Band theory to the nanoscale}
\label{Sec:2}
A \textit{photonic band} is formed by a refractive-index landscape $\epsilon(\mathbf{r}) = \epsilon(\mathbf{r}+\mathbf{a})$ , where $\mathbf{a}$ defines the photonic crystal periodicity. 
The photonic master equation is~\cite{joannopoulos1997photonic}
\begin{equation}
    \nabla \times \nabla \times \mathbf{E}(\mathbf{r}) = \frac{\omega^2}{c^2} \epsilon(\mathbf{r}) \mathbf{E}(\mathbf{r}),
    \label{master}
\end{equation}
where $\mathbf{E}(\mathbf{r})$ is the electric field with frequency $\omega$.
This equation is formally analogous to the Schrödinger equation, except that Eq.~\eqref{master} is a generalized eigenvalue problem, and has a vectorial nature compared to matter fields. 
Various numerical methods have been developed to find its solutions, including and not limited to FDTD, RCWA, T-matrix, or guided mode expansion~\cite{whittaker1999scattering, mahlau2025fdtd, asadova2025t, zanotti2024legume}. In the following, we shall focus on the latter method, to gain physical intuition on how photonic modes get coupled in the lattice.

Applying Bloch's theorem, both the electric field and the dielectric function are expanded in Fourier modes using the basis of the reciprocal lattice vectors $\mathbf{G} = (G_x, G_y)$ as $\epsilon(\mathbf{r}) = \sum_{\mathbf{G}} \epsilon_{\mathbf{G}} e^{i \mathbf{G}\cdot \mathbf{r}}$, and 
\begin{equation}
    \mathbf{E}(\mathbf{r}) = \sum_{\mathbf{G}} e^{i (\mathbf{k}+\mathbf{G})\cdot \mathbf{r}} \psi_{\mathbf{G}}(\mathbf{k}) |E_{\mathbf{G}}(\mathbf{k})\rangle.
    \label{Efield}
\end{equation} 
The state $|E_{\mathbf{G}}(\mathbf{k})\rangle$ represents the \textit{guided mode} confined in the photonic crystal, whose frequency is $\omega_\mathbf{G}=  c\sqrt{|\mathbf{k}+\mathbf{G}|}$, polarization direction is $\mathbf{p}_{\mathbf{G}}(\mathbf{k})$, and $\psi_{\mathbf{G}}$ is the coefficient to express the Bloch vector in the basis of the reciprocal lattice vectors~\cite{zanotti2022theory}. 
The $\mathbf{G}$ vectors are also known as the diffracted orders because they provide diffraction of light due to the Bragg planes of the lattice. Notice that, in this way, we have transferred all the vector nature of the electric field to the unit polarization vector $\mathbf{p}_{\mathbf{G}}(\mathbf{k})$.

The master equation is a generalized eigenvalue problem because $\hat{\epsilon}$ is not proportional to an identity matrix. The problem can be simplified to a standard eigenvalue problem by applying perturbation theory~\cite{joannopoulos1997photonic}, with the components of the effective Hamiltonian are
\begin{equation}
    \left[\hat{H}_\text{eff}(\mathbf{k})\right]_{\mathbf{G}, \mathbf{G}'} \approx \left(\omega_\mathbf{G} \delta_{\mathbf{G}, \mathbf{G}'}\right) +\mathbf{p}_{\mathbf{G}'} \cdot \mathbf{p}_\mathbf{G} \Delta \epsilon_{\mathbf{G}-\mathbf{G}'}.
    \label{diffractive}
\end{equation}

The small variation $\Delta \hat{\epsilon}$ of the dielectric function provides a coupling between waves coming from different diffracted orders, weighted by the scalar product of their polarization directions. 
Diagonalization of the effective Hamiltonian in Eq.~\eqref{diffractive} $\hat{H}_\text{eff}(\mathbf{k}) |\psi(\mathbf{k})\rangle  = \omega(\mathbf{k}) |\psi(\mathbf{k})\rangle $ yields the bands and the modes $|\psi(\mathbf{k})\rangle$ of the photonic crystal. 

As we shall see in the following, the properties of the bulk band eigenmodes play a fundamental role in defining the \textit{bulk topology}.
However, it is impractical in experiments to have access to these bulk eigenmodes. Most optical characterization relies on observing the light that leaks in the far field, which will require defining the properties of photonic crystals from non-Hermitian principles, as we will see next.

\subsection{Topological band theory in the bulk}
Geometrical and topological quantities, such as the Berry curvature or Berry phase, have been introduced in the bulk of photonics models in complete analogy to their solid-state counterparts~\cite{wang2008reflection, blanco2020tutorial}. 
From the eigenfunction $|\psi(\mathbf{k})\rangle$ of the effective Hamiltonian in Eq.~\eqref{diffractive}, the Berry connection is
\begin{equation}
    \mathbf{A}(\mathbf{k}) = i \langle \psi(\mathbf{k}) | \nabla_\mathbf{k} |\psi(\mathbf{k})\rangle.
    \label{Ab}
\end{equation}
The integral of the connection on a closed path $\mathcal{C}$ in the parameter space yields the Berry phase
\begin{equation}
    \varphi_B = \oint_\mathcal{C} \mathbf{A}(\mathbf{k}) \cdot d\mathbf{k}.
    \label{Bpb}
\end{equation}
Oftentimes, photonic crystals present energy degeneracy, and bands are clustered together without band gaps, requiring the use of Wilson loops defined from a multiband Berry connection, where the bra-ket product in Eq.~\eqref{Ab} is evaluated with the connected bands~\cite{blanco2020tutorial}.

The Berry connection is a gauge-dependent quantity, i.e., it varies when the wavefunction is varied by a global phase factor. However, the Berry phase is gauge invariant up to multiples of $2\pi$.
Taking the curl of Eq.~\eqref{Ab} yields another gauge-invariant quantity, namely the Berry curvature
\begin{equation}
    \mathbf{B}(\mathbf{k}) = \nabla_\mathbf{k} \times \mathbf{A}(\mathbf{k}).
    \label{Bb}
\end{equation}
For 2D systems, the only nonzero component of the Berry curvature is $B_z$, and its integral over the entire Brillouin zone is the Chern number $C=\frac{1}{2\pi}\int_{\text{BZ}} B_z d^2k$ a topological invariant of the integer quantum Hall effect. 
The Berry curvature is the imaginary antisymmetric part of the so-called quantum geometric tensor~\cite{kolodrubetz2017geometry}
\begin{equation}
    T_{\mu\nu} = \langle \partial_\mu \psi| \partial_\nu \psi\rangle - \langle \partial_\mu \psi | \psi \rangle \langle \psi | \partial_\nu \psi\rangle 
\end{equation}
where $\partial_\mu$ defines the derivative along the direction of $\mathbf{k}_\mu$, with $\mu=\lbrace k_x, k_y\rbrace$, such that 
\begin{equation}
    B = -2\mathrm{Im}\lbrace T_{k_x, k_y} \rbrace, \quad g_{\mu\nu} =\mathrm{Re}\lbrace T_{\mu, \nu} \rbrace.
\end{equation}
The real part of the geometric tensor defines the distance between quantum states in parameter space and is known as the \textit{quantum metric}. Whereas Berry curvature has historically played a pivotal role in topological photonics effects, recent works have shown that quantum metric becomes a crucial quantity that can potentially appear in many physical phenomena, from anomalous Hall effect and wavepacket dynamics in polariton systems~\cite{Gianfrate2020, hu2024generalized, hu2025quantum}, to exceptional points in non-Hermitian systems~\cite{solnyshkov2021quantum, liao2021experimental}.

\subsection{Non-Hermitian systems}
Recent years have seen a growing interest in the interplay of topology and non-Hermiticity~\cite{parto2020non, Monticone, BergholtzRev, ding2022non, wang2023non, wang2021topologicalreview, shen2018topological}. Non-Hermitian photonics was first introduced to describe parity-time-symmetric systems, including systems with optical loss and gain, such as lasers and materials with strong absorption~\cite{OtaActiveTopoPhoton, li2023exceptional, parto2020non}.

The most obvious consequence of non-Hermitian systems is that their Hamiltonians admit, in general, complex-valued eigenenergies where the imaginary part is related to losses. Due to this complex nature, bands in non-Hermitian systems may possess richer properties compared to their Hermitian counterparts~\cite{torres2019perspective, wang2023non}. 
For example, band gaps require careful definition and are generally classified into two main categories. Line gaps arise when the spectrum avoids crossing a reference line, e.g., the real axis, similarly to conventional Hermitian gaps. Point gaps exist when the eigenvalues avoid a chosen reference point in the complex plane. These latter gaps support a unique form of \textit{spectral topology}, in which the eigenvalues can braid over the Brillouin zone, forming topologically nontrivial loops in the complex energy plane as a function of momentum~\cite{wang2021topological, wang2021generating, wojcik2022eigenvalue}.

Moreover, band singularities in non-Hermitian models can lead to diabolic points, analogous to the Hermitian counterparts, but also to exceptional points (EPs), which form when the Hamiltonian becomes nondiagonalizable. At EPs, the complex bands experience a branch cut, and eigenvectors coalesce because they are no longer orthogonal~\cite{ozdemir2019parity, li2023exceptional, ding2022non, miri2019exceptional}. EPs are extremely interesting for applications such as telecommunication or sensing, due to the enhanced sensitivity to mode switching~\cite{meng2024exceptional}.

While band gaps and singularities describe the features of non-Hermitian systems with periodic boundary conditions, open boundary conditions also lead to new phenomena that break down the usual bulk-boundary correspondence. Unlike in Hermitian models, the non-Hermitian eigenstates can become exponentially localized at the boundaries of the system rather than being extended, due to a failure of the Bloch description. This effect, known as the skin effect, originates from the impossibility of forming standing waves with opposite momenta and is typically observed in systems with nonreciprocity. To solve this issue, a generalized Brillouin zone is introduced, where momentum is complex, defining a non-Bloch theory~\cite{yao2018non, yang2020non, yokomizo2019non, yokomizo2023non}. The emergence of the skin effect has been related to the spectral topology of the bands~\cite{wang2024non, okuma2020topological, zhang2020correspondence, zhong2021nontrivial, zhang2022universal}, also in connection with EPs in 2D photonic crystals~\cite{fang2022geometry, zhong2021nontrivial}. 

In addition to spectral topology, EPs, and skin effect, nontrivial band topology can still arise in analogy to that of Hermitian systems, and a full classification of non-Hermitian band topology has been systematically extended across symmetry classes~\cite{UedaRev, zhou2019periodic, kawabata2019symmetry}. We now outline how Berry curvatures are properly defined in the non-Hermitian framework.

\subsubsection{Berry curvature in non-Hermitian systems}
When the non-Hermitian Hamiltonian can be diagonalized both from the left and the right, it yields left and right eigenvectors. These eigenvectors $|\psi^R\rangle \neq |\psi^L\rangle$ share the same eigenvalue 
\begin{equation}
\hat{H}|\psi^R\rangle= \varepsilon \;|\psi^R\rangle; \qquad \quad
\langle\psi^L|\hat{H}= \langle\psi^L|\varepsilon.    
\end{equation}
Although the norm of these states is not conserved due to the losses or gains, leading to intensity amplification, one usually imposes that $\langle\psi^L_i|\psi^R_j\rangle = \delta_{ij}$, thus requiring the states to be bi-orthonormalized. 

With a combination of the left and right eigenvectors and Eq.~\eqref{Bb}, the following non-Hermitian Berry curvature can be obtained~\cite{zhang2019quantum}
\begin{equation}
    B^{\alpha \beta} = i \left(\langle \partial_{\mu} \psi^{\alpha} | \partial_{\nu} \psi^{\beta}\rangle -\langle \partial_{\nu} \psi^{\alpha} | \partial_{\mu} \psi^{\beta}\rangle\right),
    \label{eq:qgt}
\end{equation}
for $\alpha, \beta =L,R$, so that four locally different Berry curvatures can be obtained ($B^{RR}$, $B^{LR}$, $B^{RL}$, $B^{LL}$), but all of them integrate to the same Chern number~\cite{shen2018topological}. While the $RR$ Berry curvature is a real number, the $LR$ and $RL$ curvatures admit, in general, an imaginary part, which has been related to the intensity amplification factor and the quantum Hall susceptance~\cite{fan2020complex, ozawa2025geometric}. 

Both right and left eigenvectors are needed for calculating the full non-Hermitian topological properties. On the one hand, the former represent the physical field distributions radiating into the far field, and are easily experimentally accessible, but the latter, on the other hand, are not. In two-band polariton experiments, the left eigenvectors have been reconstructed from the Stokes parameters of the right states~\cite{hu2024generalized}. However, most nanophotonic crystals are inherently multiband; thus, the interest has been concentrated on the right eigenvectors only, failing to recover the left eigenvectors. 

Finding good observables that can return the left eigenvectors is important for deepening the understanding of non-Hermitian topology. For example, in mechanical metamaterials with nonreciprocity, left states have been linked to the system's response to external perturbations~\cite{schomerus2020nonreciprocal}. This suggests that it may be possible, also in nanophotonics, to reconstruct the spatial profiles of left eigenstates by applying small perturbations or phase-sensitive imaging techniques, thereby providing deeper insights into the full non-Hermitian properties of the system. 

\subsubsection{BICs from an effective non-Hermitian model}
Guided resonances in photonic crystals generally experience radiation losses. However, bound states in the continuum (BICs) represent special modes within these resonances where such losses are suppressed due to symmetry-driven interference mechanisms. This motivates the development of a non-Hermitian model starting from Eq.~\eqref{diffractive} to capture the lossy nature of the photonic crystals and define the modes' lifetime or Q-factor. 

In the following, we present a simple toy model that considers lossless guided modes coupled to a continuum of lossy non-guided modes above the light line. Various mechanisms, including the existence of boundaries, can facilitate such coupling and create a non-radiating and absorption-free BIC. ~\cite{kolkowski2023enabling}.

For the sake of simplicity, consider two guided modes $|1\rangle$ and $|2\rangle$ with the same polarization, and frequencies $\omega_{1,2}$ respectively. These modes are coupled with strength $\Omega$, yielding the Hamiltonian $\hat{H}_{12} = \sum_{j=1,2} \omega_j |j\rangle\langle j | + (\Omega |2\rangle\langle1| + \text{H.c.})$, similarly to Eq.~\eqref{diffractive}.
We now introduce a third mode, the continuum state $|c\rangle$, with frequency $\omega_c = 0$ without loss of generality. The interaction between the guided modes and the continuum is described by Lindblad operators $\hat{L}_{j} = \sqrt{\gamma_r} |c\rangle\langle j| - \sqrt{\gamma_r} |j\rangle\langle c|$, for $j=1,2$, capturing both radiative loss and back-action into the guided modes.
The Lindblad master equation for the density matrix $\rho$ can be written as $\dot{\hat{\rho}} = -i \left[\hat{H}_{12}^\text{eff}, \hat{\rho}\right] + \sum_j \hat{L}_j \hat{\rho}\hat{L}_j^\dagger$, where we have rearranged terms to define an effective non-Hermitian Hamiltonian~\cite{roccati2022non}
\begin{equation}
    \hat{H}_{12}^\text{eff} = \hat{H}_{12} + \frac i 2 \sum_j \hat{L}^\dagger_j \hat{L}_j.
    \label{eq:hamiltonian_effective}
\end{equation}
Since we are only interested in the guided modes, we project onto the $|1\rangle$ and $|2\rangle$ modes and neglect the jump operator's contribution. Given the explicit form of $\hat{L}_j$, the second term  Eq.~\eqref{eq:hamiltonian_effective} is the loss term 
\begin{equation}
    \hat{\gamma} =  \frac i 2 \sum_j \hat{L}^\dagger_j \hat{L}_j = - i \gamma_r \begin{pmatrix}
        \quad\!1 & -1 \\
        -1 & \quad\!1
    \end{pmatrix},
\end{equation}
whose diagonal part represents the guided mode intrinsic losses, while the off-diagonal part is the radiative coupling between the co-polarized guided modes. 

In general, we consider the set of guided modes spanned by the diffracted orders $\mathbf{G}$ as introduced before, each with its polarization vector $\mathbf{p}_{\mathbf{G}}$. Then, the radiative coupling between two such modes depends on the angle formed by their polarization vectors, and the effective Hamiltonian in Eq.~\eqref{diffractive} becomes
\begin{equation}
    \left[\hat{H}_\text{eff}^{\text{rad}}(\mathbf{k})\right]_{\mathbf{G}, \mathbf{G}'} \approx \left(\omega_\mathbf{G} \delta_{\mathbf{G}, \mathbf{G}'}\right) + \left( \Delta \epsilon_{\mathbf{G}-\mathbf{G}'} - i \gamma_r \right) \mathbf{p}_{\mathbf{G}'} \cdot \mathbf{p}_\mathbf{G} 
    \label{Htot}
\end{equation}
Crucially, only the Fourier components corresponding to diffracted orders that fall within the light line are observed, so the experimentally accessible signal is inherently a projection of the bulk eigenmode onto a finite number $N$ of channels, spanned by the diffracted order vectors $\mathbf{G}$. The electric field in the far field can be defined as~\cite{yin2024observation, yuan2025breakdown}
\begin{equation}
\mathbf{E}_\text{F} \propto \hat{R} |\psi\rangle = \sum_\mathbf{G} c_\mathbf{G} \mathbf{p}_\mathbf{G},
\label{farfield}
\end{equation}
where $\hat{R}$ is a $N\times 2$ matrix containing $N$ (2 components) column vectors of the polarization directions $\mathbf{p}_\mathbf{G}$, and $|\psi\rangle = \left(\dots c_\mathbf{G} \dots \right)^T$ is the $N$ components eigenmode of the effective Hamiltonian in Eq.~\eqref{Htot}. Notice that $\hat{R}$ is not a unitary transformation, and only admits a pseudo-inverse, such that $\hat{R} \hat{R}^{-1} = \mathbb{I}_2$ is a $2\times 2$ identity matrix, while $\hat{R}^{-1} \hat{R} \neq \mathbb{I}_2$.

The non-Hermitian Hamiltonian description in Eq.~\eqref{Htot} was introduced to explain the formation of symmetry-protected BIC in exciton-polariton systems on 1D gratings~\cite{ArdizzoneBIC, lu2020engineering, sigurdhsson2024dirac}. Within this description, one can show that the far-field expression in Eq.\eqref{farfield} leads to a polarization vortex for BICs, i.e., states with arbitrarily small radiative losses. While polarization singularities may arise even in Hermitian systems, the inclusion of non-Hermitian effects allows for a formal connection between these topological singularities and BICs with high Q-factors. This formalism was also recently extended to 2D structures~\cite{nguyen2025generalized}, where the interplay between radiative coupling and diffractive coupling of guided modes leads to the formation of a BIC flat band~\cite{do2025room}, tunable Friedrich–Wintgen BICs at a highly oblique angle~\cite{mermet2023taming}, and Berry curvature with non-Hermitian origin~\cite{yuan2025breakdown}. Building on this framework, we now discuss what such a non-Hermitian Hamiltonian reveals about \textit{far-field topology} in photonic systems. 

\subsection{Band topology in the far-field}
\label{sec:farfield}
So far, we have obtained the eigenstates of the bulk effective non-Hermitian Hamiltonian, including radiation losses. We have then connected such states to the corresponding electric far field, in Eq.~\eqref{farfield}. The electric far field can be further rewritten using the circular polarization basis
\begin{equation}
    |E \rangle = \cos(\theta/2) e^{i \phi} |+\rangle + \sin(\theta/2) |-\rangle,
    \label{farfieldcp}
\end{equation} 
where the angles parameterize the state on the Poincaré sphere, representing the polarization state of light. This state can also be expressed using the Stokes parameters $S_i = \langle E | \sigma_i | E \rangle $, where $\sigma_i$ is the $i$-th Pauli matrix. It is straightforward to verify that $S_z = \cos\theta$, $S_y= - \sin\theta \sin\phi$, $S_x= \sin\theta \cos\phi$. Such parametrization is useful for experiments, where the Stokes parameters are obtained from the intensities measured in each of the six polarizations of light.

Using the state in Eq.~\eqref{farfieldcp}, the far-field Berry connection is defined analogously to the bulk Berry connection in Eq.~\eqref{Ab} as
\begin{equation}
    \mathbf{A}_f(k_x, k_y) = i \langle E | \nabla_\mathbf{k} | E \rangle = - \cos^2\frac\theta 2 \left(\partial_{k_x} \phi \hat{k}_x + \partial_{k_y} \phi \hat{k}_y \right)
\end{equation}
where $\hat{k}_\mu$ defines the unit direction of a vector along $\mathbf{k}_\mu$.
The corresponding far-field Berry curvature, from Eq.~\eqref{Bb} using Eq.~\eqref{farfieldcp}, is
\begin{equation}
\begin{split}
    B_f(k_x,k_y) &= \nabla_\mathbf{k} \times \mathbf{A}_f(k_x, k_y) \cdot \hat{k}_z\\&= \frac 1 2 \sin\theta \left(\partial_{k_x}\theta \partial_{k_y}\phi - \partial_{k_y}\theta \partial_{k_x}\phi\right)
\end{split}
\end{equation}
This formula has been widely used to extract the Berry curvature in exciton-polariton planar microcavities~\cite{bleu2018measuring, polimeno2021tuning, Gianfrate2020, hu2024generalized}, liquid crystal cavities~\cite{lempicka2022electrically}, and in plasmonic nanoparticle arrays reduced to two-band models~\cite{yin2024observation, cuerda2024pseudospin, cuerda2024observation}.

The Berry phase, analogous to Eq.~\eqref{Bpb}, using Eq.~\eqref{farfieldcp} can be calculated as
\begin{equation}
\begin{split}
    \varphi_f &= \oint_\mathcal{C} \mathbf{A}_f \cdot d\mathbf{k} = -\frac 1 2 \oint_\mathcal{C} \left(1-\cos\theta\right) \nabla_\mathbf{k} \phi \cdot d \mathbf{k} \\&= - \pi q + \frac{1}{2}\oint_\mathcal{C} \cos\theta \nabla_\mathbf{k} \phi \cdot d \mathbf{k} =  - \pi q +\varphi_G
    \label{berryphaseff}
\end{split}
\end{equation}
having also used the definition in Eq.\eqref{topocharge_def} to recover the topological charge $q$. The path $\mathcal{C}$ denotes a closed trajectory in the Brillouin zone, along which each eigenstate is mapped with a polarization state on the Poincaré sphere. This defines a corresponding path on the sphere, whose shape depends on the specific polarization texture of the band. The second term $\varphi_G = \frac{1}{2}\oint_\mathcal{C} \cos\theta \nabla_\mathbf{k} \phi \cdot d \mathbf{k}$ is a geometrical phase equal to the area between this path and the equator on the Poincaré sphere~\cite{bliokh2019geometric}. 
This formula means that a polarization vortex of charge $q$ is a source of $-\pi q$ Berry phase, a feature that has been utilized to obtain optical spin-Hall effect from the topological charge of BICs~\cite{wang2022spin, wan2025photonic}. 
Such interplay between non-Hermitian physics and far-field observables makes photonic crystals suited for exploring novel topology beyond the standard bulk-band correspondence.

\section{Limitations and future directions}
\textit{Bulk and far field topologies} have typically been studied independently, but it is important to understand if and under what conditions one can be inferred from the other. This issue is particularly relevant in the presence of BICs, where information from the light wavefunction might be lost in the far-field interference. 
Topological charges of BICs can be exchanged during band inversions, suggesting they can be a good indicator for topological phase transitions~\cite{Salerno2022, bouteyre2022non, ji2024probing, zhang2025topology}. However, these studies focused on the local properties of BICs in momentum space, failing to recover information from the full Brillouin zone, which is supposed to yield the correct topological invariant.

A symmetry representation of field eigenmodes at high-symmetry points of the Brillouin zone is, in principle, enough to characterize the topology of photonic crystals~\cite{vaidya2023topological, fang2012bulk}, which is typically where BICs and far field polarization vortexes appear. This suggests that the global topology of a band in the far field can indeed be estimated by the total number of polarization singularities, both vortexes of linearly- and circularly-polarized states~\cite{fosel2017lines, arjas2025}.

Recently, scientists have studied the correspondence between the bulk and far field topology, finding that it might break down~\cite{yin2024observation, yuan2025breakdown}. Namely, while far-field radiation provides experimental access to topological features, it also modifies observables in ways that are not fully captured by their corresponding bulk ones. The Berry curvature and the Berry phase are altered due to the presence of polarization singularity points.
This correspondence becomes particularly significant when the Berry curvature is directly tied to measurable effects, such as anomalous transport or other optical responses. In such cases, the distinction between bulk and far-field Berry curvature might become relevant not only because it may affect quantitative predictions and the interpretation of experimental data, but also for understanding how radiation alters intrinsic topological properties.

This discrepancy between bulk and far-field topology suggests a broader issue about photonic systems operating in regimes that fall outside the assumptions of standard topological band theory. As we have seen in Sec.~\ref {Sec:2}, photonic systems are described by a generalized eigenvalue problem, often in combination with nonlinear and non-Hermitian effects. An effective Hamiltonian description and, hence, band theory can be recovered in some regimes. However, these generalized eigenvalue problems are, in principle, beyond the standard topological band theory. The field is still in its early stage, where scientists are now studying ideal observables to directly map topological phenomena in non-Hermitian multi-band systems~\cite{yin2024observation, yuan2025breakdown}, with potential applications to hyperbolic metamaterials or negative index photonic crystals~\cite{isobe2025topological}. Moreover, the study of intrinsic material anisotropy might lead to novel Dirac-like degeneracies unique to photonic bands~\cite{xiong2020hidden, de2022manipulating}.

Looking ahead, promising directions include the use of structured light and interferometry to probe polarization vortices, and the reconstruction of non-Hermitian topological invariants from radiation patterns. This ongoing effort to generalize topological band theory in nanophotonics naturally extends to systems with explicit time dependence, where temporal modulation of system parameters is used to achieve topological phenomena~\cite{galiffi2022photonics}.
\textit{Floquet theory} is a well-established framework for treating quantum
system under periodic time modulations, used to induce properties that are absent in their static counterpart. Floquet topological insulators so far made in photonics are based on the paraxial approximation, which exploits the analogy between time and the direction in which light propagates~\cite{rechtsman_photonic_2013, mukherjee_experimental_2017, maczewsky_observation_2017}. Inspired by these works, scientists have realized spatially modulated metallic nanowires that can act as analogs of optical fiber to guide and evanescently couple surface plasmon polariton modes with a nanometer scale and realize a Floquet plasmonic topological insulator, robust to the intrinsic loss of plasmonic modes~\cite{zhang2017plasmonic}. A drawback of such an approach is that the realized system requires a relatively long third direction, of the order of 10 times the wavelength, along which the nanowires are modulated. Such length is impractical for nanoscale on-chip applications.
A possible solution could be to apply a Floquet parametric drive at optical frequencies to the plasmons, as done for solids to enhance and control the resonance and create exceptional points~\cite{kiselev2024inducing}. This approach could be used to design a 2D lattice of nanoparticles where each particle's surface plasmon resonance is dynamically modulated, enabling new collective states and tunable optical properties.

Another powerful framework that has been widely used to implement nontrivial topology is \textit{synthetic dimensions}~\cite{OzawaPrice_Rev}. In this approach, some internal degrees of freedom of the system are interpreted as sites along a fictitious spatial direction, expanding a system’s dimensionality, with motion between sites enabled by an external force. Bringing synthetic dimensions to the nanoscale might offer even greater advantages, such as enhanced field confinement, access to otherwise inaccessible topological regimes, or even create nonstandard hyperbolic or fractal geometries~\cite{grass2025colloquium}.
Realising one or more synthetic dimensions, in addition to the real ones, could allow for unprecedented explorations of non-Hermitian topological effects in three or four-dimensional nanophotonic systems. Moreover, combining synthetic dimension with nearest-neighbor coupling and a real spatial dimension with long-range diffractive coupling can lead to hybrid dimensional physics with engineered connectivity, extending the current possibilities to explore an even larger parameter space for topological protection in open systems. 

One of the candidates for the implementation of such a synthetic dimension could be represented by the \textit{topological rainbows}~\cite{lu2021topological}, a class of photonic crystals in which a system parameter, e.g., lattice spacing or site size, varies smoothly in space, creating an interface direction. As a result, different frequencies of light separate into several positions due to topological properties of the system, akin to colors of a rainbow~\cite{elshahat2021perspective, lu2022chip}. If these frequency components are coherently coupled, the setup can simulate motion along the synthetic dimension, with frequency acting as an extra spatial coordinate~\cite{SegevSynthetic}.

To further enrich this picture, spatial variation of the photonic lattice parameters in a third (real or synthetic) dimension can also enable the realization of 3D non-trivial topology, as well as Weyl points and Fermi arcs~\cite{grossi2023quasicrystalline, nguyen2023fermi}. 
Stacking layers of photonic lattices along a real spatial axis has been shown to introduce long-range coupling in the far field, and the possibility to realize moiré structures when the layers are slightly twisted or mismatched in periodicity~\cite{guan2023far}.

\textit{Moiré photonic crystals} are another powerful system that is emerging for observing flat bands associated with reduced group velocity and enhanced light–matter interaction~\cite{nguyen2022magic, jing2025observation, saadi2025tailoring}, as well as for realizing nanolasers operating at a magic angle, exhibiting both high spatial and spectral coherence~\cite{mao2021magic}. Additionally, these moiré structures allow for the exploration of BICs~\cite{huang2022moire, qin2024optical} or skyrmion phenomena~\cite{schwab2025skyrmion}, optical vortex generation~\cite{zhang2023twisted}. 

The integration of Floquet modulations with synthetic dimensions and moiré-engineered structures hosting polarization singularities could pave the way for the next generation of topological photonic devices and advanced light manipulation.

\subsection*{Concluding remarks}
Topological photonics is a powerful framework that has matured rapidly over the past decade, both in fundamental science and technological applications. Yet, translating these concepts to the nanoscale brings new challenges that lie outside the scope of conventional topological models. 

Far-field radiation may be a rich source of topological structure and nonlocal effects, manifesting through BICs or circularly-polarized states. The far-field observables, however, do not always correspond directly to bulk topological properties and go beyond the conventional definition of topology. Theoretical approaches that incorporate generalized eigenvalue problems, full non-Hermitian Berry curvatures, and response functions from left eigenvectors may be fundamental for capturing the full topological picture. Ultimately, pushing topological photonics into the strong coupling regime might unlock entirely new classes of topological quasiparticles and robust quantum optical phenomena in the nanoscale domain.

\begin{acknowledgments}
Thanks to K. Arjas, A. Gianfrate, H.S. Nguyen, and D. Trypogeorgos for fruitful discussions. This work was supported by the Research Council of Finland under the Academy Research Fellowship project n.~13354165, and by the Italian Ministry of University and Research under the Rita Levi-Montalcini program.
\end{acknowledgments}
Data sharing is not applicable to this article as no new data were created or analyzed in this study.
\providecommand{\noopsort}[1]{}\providecommand{\singleletter}[1]{#1}%

\end{document}